\documentclass[pre,aps,twocolumn,superscriptaddress]{revtex4-1}
\pdfoutput=1

\usepackage{amssymb}
\usepackage{epsfig}
\usepackage{amsmath}
\usepackage{subfigure}
\usepackage{graphicx}
\usepackage{textcomp}
\usepackage{url}
\usepackage{float}
\usepackage{color}
\usepackage{cases}
\usepackage[normalem]{ulem}

\usepackage[titletoc,title]{appendix}

\usepackage[colorlinks=true, urlcolor=blue, anchorcolor=blue, citecolor=blue,filecolor=blue,linkcolor=blue,menucolor=blue
]{hyperref}

\usepackage{color}

\begin{document}
\title{Persistent fluctuations of the swarm size of Brownian bees}

\author{Baruch Meerson}
\email{meerson@mail.huji.ac.il}
\affiliation{Racah Institute of Physics, Hebrew University of
Jerusalem, Jerusalem 91904, Israel}
\author{Pavel Sasorov}
\email{pavel.sasorov@gmail.com}
\affiliation{Institute of Physics CAS, ELI Beamlines, 182 21 Prague, Czech Republic}
\affiliation{Keldysh Institute of Applied Mathematics, Moscow 125047, Russia}

\begin{abstract}
The ``Brownian bees" model describes a
system of $N$ independent branching Brownian particles. At each branching event the particle farthest
from the origin is removed, so that the number of particles remains constant at all times. Berestycki \textit{et al.} (2020) 
proved that, at $N\to \infty$,  the coarse-grained spatial density of this particle system lives in a spherically symmetric domain and is described by
the solution of a free boundary problem  for a deterministic reaction-diffusion equation. Further, they showed that, at long times, this solution approaches a unique spherically symmetric steady state with compact support: a sphere which radius $\ell_0$ depends on the spatial dimension $d$. Here we study fluctuations in this system in the limit of large $N$ due to the stochastic character of the branching Brownian motion, and we focus on persistent fluctuations of the swarm size.  We
evaluate the probability density $\mathcal{P}(\ell,N,T)$ that the maximum distance of a particle from the origin
remains smaller than a specified value $\ell<\ell_0$, or larger than a specified value $\ell>\ell_0$,  on a time interval $0<t<T$, where $T$ is very large. We argue that $\mathcal{P}(\ell,N,T)$  exhibits the large-deviation form $-\ln \mathcal{P} \simeq N T R_d(\ell)$.  For all $d$ we obtain asymptotics of the rate function $R_d(\ell)$ in the regimes $\ell \ll \ell_0$, $\ell\gg \ell_0$ and $|\ell-\ell_0|\ll \ell_0$. For $d=1$ the whole rate function can be calculated analytically.  We obtain these results by determining the optimal (most probable) density profile of the swarm, conditioned on the specified $\ell$, and by arguing that this density profile is spherically symmetric with its center at the origin.
\end{abstract}

\maketitle
\nopagebreak

\section{Introduction}
\label{intro}

Nonequilibrium steady states of macroscopic systems composed of reacting and diffusing particles, continue to
attract attention from physicists \cite{JL1993,JL2004,Bodineau2010,Hurtado2013,M2015}. The nonequilibrium steady states shed light
on the physics of a plethora of important dissipative systems, both non-living and living. The search for, and the analysis of, simple models which can teach us about general properties of nonequilibrium fluctuations, continues. One of the simplest models of this type is a particle-conserving variant of branching Brownian motion that we will describe shortly.

Branching Brownian motion (BBM for short) unites two fundamental continuous-time Markov processes: the random branching and
the Brownian motion, or the Wiener process. In the past BBM was extensively studied by mathematicians \cite{McKean,Bramson}, and it continues to attract interest from physicists \cite{BD2009,Mueller,Ramola1,DMS}. Here we will consider the recently formulated \emph{Brownian bees} model \cite{bees1,bees2}.

``Brownian bees" is a 
variant of BBM with an imposed exact conservation law. The microscopic model is
defined as follows. The system consists of $N$ independent particles (bees) located in a $d$-dimensional space.
Each particle  can branch, in a small
time interval $\Delta t$,  into two particles
with probability $\Delta t$. It can also perform, with the complementary probability $1-\Delta t$, continuous-time Brownian motion with diffusion constant $1$ \cite{rescaling}. Whenever a branching event occurs, the particle which is farthest from
the origin is instantaneously removed, so that the number of particles remains constant at all times.  The name Brownian bees, coined by J. Quastel \cite{bees1}, comes from the
superficial analogy with a swarm of bees around a hive.

The Brownian bees is a particular example of the so called Brunet-Derrida $N$-particle systems. All these systems
involve the BBM with exact conservation law. They differ from each other only by the rule of elimination of ``least-fit" particles, thus providing insight into different aspects of  biological selection. See Ref. \cite{bees2} for a brief review of these models.

Recently Berestycki \textit{et al.} \cite{bees1}
showed that, in the limit of $N\to \infty$, the coarse-grained spatial density $u(\mathbf{x},t)\geq 0$ of the Brownian bees is described by
the solution of the following deterministic free boundary problem in $d$ dimensions:
\begin{eqnarray}
\label{deterministicd}
  &&\partial_t u (\mathbf{x},t) = \nabla^2 u(\mathbf{x},t)+u(\mathbf{x},t)\,,\quad |\mathbf{x}|\leq L(t)\,, \nonumber\\
  &&u(\mathbf{x},t)= 0\,, \quad |\mathbf{x}|> L(t)\,, \nonumber \\
  &&\int \displaylimits_{|\mathbf{x}|<L(t)}  u(\mathbf{x},t) \,d\mathbf{x}= 1\,.
\end{eqnarray}
$u(\mathbf{x},t)$ is continuous at $|\mathbf{x}|=L(t)$, and an initial condition must be specified. According to
Eq.~(\ref{deterministicd}), $u(\mathbf{x},t)$ has a compact support which is, at all $t>0$, a $d$-dimensional sphere centered at the origin. There is an effective absorbing wall at $|\mathbf{x}|=L(t)$, which moves so as to impose the constant number of particles at all times.

In a companion paper \cite{bees2}  Berestycki \textit{et al.}
showed that, at long times, the solution of the deterministic problem~(\ref{deterministicd}) approaches
a unique steady state $u=U(\mathbf{x})$ which is described by the  fundamental mode of the Helmholtz equation with a unit eigenvalue,
\begin{equation}\label{Helmholtz}
 \nabla^2 U+U = 0\,,
\end{equation}
inside a $d$-dimensional sphere, whose center is at the origin, and whose  radius $\ell_0$ depends on $d$. The steady state density profile is spherically symmetric  and has the following form:
\begin{numcases}
{{U(r)} =}\frac{F(r)}{4 \pi \ell_0^{d/2}
   J_{\frac{d}{2}}(\ell_0)}\,, & $r<\ell_0$, \label{Ur} \\
0\,,& $r>\ell_0$. \label{outside}
\end{numcases}
where
\begin{equation}\label{F}
F(r) = \frac{J_{\frac{d}{2}-1}(r)}{r^{\frac{d}{2}-1}}\,,
\end{equation}
and $\ell_0=\ell_0(d)$ is the first positive root of the Bessel function of the first kind $J_{d/2-1}(r)$.
For $d=1$ one obtains
\begin{numcases}
{{U(x)} =}\frac{1}{2}\cos x\,, & $|x|<\ell_0$, \label{Ux} \\
0\,,& $|x|>\ell_0$. \label{outsidex}
\end{numcases}
and $\ell_0=\pi/2$. As the solution of the deterministic free boundary problem~(\ref{deterministicd}) approaches the steady state, $L(t)$ approaches $\ell_0$.

In this work we study \emph{fluctuations} of a stationary swarm of Brownian bees due to the stochastic character of the branching Brownian motion. We consider the limit of $N\gg 1$. In this limit the fluctuations are  typically small. But large fluctuations (often called large deviations) also occur, and it is interesting to evaluate their probability as well. We focus on \emph{persistent} fluctuations of the maximum distance $L(t)$ of a bee from the origin.  Our objective is to
evaluate the probability  that, on a \emph{long} time interval $0<t<T$,  $L(t)$ remains smaller than a specified value $\ell<\ell_0$, or larger than a specified value $\ell>\ell_0$.  We argue that,  at $N\gg 1$ and $T \to \infty$, the corresponding probability density $\mathcal{P}(\ell,N,T)$ exhibits the large-deviation form
\begin{equation}\label{LDform}
-\ln \mathcal{P}(\ell,N,T) \simeq N T R_d(\ell)\,.
\end{equation}
This result follows naturally from the optimal fluctuation method (OFM) which we employ for solving this problem.
The OFM  (also known in other fields as the instanton method, the weak noise theory and the macroscopic fluctuation theory) is briefly described
in Sec.~\ref{governing}. The OFM boils down to finding the optimal (most probable) density profile of the swarm, conditioned on the specified $\ell$ during a long time $T$, which dominates the probability density of $\ell$.

Our result~(\ref{LDform}) relies on two assumptions. First, we assume that, at large $T$, the optimal density profile has a \emph{spherical} compact support $L(t)$. We will discuss this assumption in Sec. \ref{summary}.  (Recall that a spherical compact support is also observed, at any $t>0$, for the most probable (unconditioned) density profile of the Brownian bees when $N\to\infty$~\cite{bees1}.) Second, we assume that, to  leading order in $T\gg 1$,  the optimal density profile is spherically symmetric and independent of time, whereas the support radius $L(t)$ is equal to $\ell$.

Based on these assumptions, we  derive, in Sec. \ref{threeasymp}, analytical asymptotics of the rate function for all $d$ in the regimes $|\ell-\ell_0|\ll \ell_0$, $\ell \ll \ell_0$ and $\ell\gg \ell_0$. These asymptotics have the following form:
\begin{numcases}
{{R_d(\ell)} \simeq}\alpha_d (\ell-\ell_0)^2\,,& $|\ell-\ell_0|\ll \ell_0$, \label{wnlineartheory}\\
\frac{\ell_0^2}{\ell^2} - \frac{2}{\sqrt{\alpha_d} \,\ell} & $\ell\ll\ell_0$\,, \label{ellsmall} \\
\frac{1}{3}\,,& $\ell\gg\ell_0$\,,\label{elllarge}
\end{numcases}
where the $d$-dependent factor $\alpha_d$ is presented in Eq.~(\ref{alphad}) below. For $d=1$ the whole rate function $R_d(\ell)$ can be calculated exactly, as we explain in Sec. \ref{1dcase}. We summarize and briefly discuss our results in Sec. \ref{summary}.

\section{Optimal fluctuation method: Governing equations}
\label{governing}

A convenient departure point for the derivation of the OFM equation is a lattice gas formulation for a gas of non-interacting branching random walkers. One starts from a multivariate master equation which describes the evolution with time of the probability of observing a certain number of particles on each lattice site at time $t$. Being interested in large deviations, and making a WKB-type ansatz in the master equation,  one arrives at an effective multi-particle Hamilton-Jacobi equation, which can be recast in a Hamiltonian form \cite{EK,MS,MSK}.   Assuming in addition that the hopping rate of the random walkers is much larger than the branching rate, one obtains a continuous coarse-grained Hamiltonian field-theoretic description of large fluctuations in this reacting lattice gas, valid at distances large in comparison with the lattice constant \cite{EK,MS,MSK}. In this limit the lattice constant only enters (alongside with the hopping rate) the diffusion constant, bringing back the continuous-space BBM model.
In addition to the gas density field $q(\textbf{x},t)$, which formally plays the role of the ``coordinate" of the Hamiltonian description,  there is a canonically conjugate ``momentum" density field $p(\textbf{x},t)$ which describes
the most likely configuration of the noise which dominates the large deviation in question.

The OFM equations for reacting lattice gases were also derived and used by other workers \cite{MFT,JL1993,JL2004,Bodineau2010,MSFKPP,MVS}.
The problem of Brownian bees, however, brings an important new element: exact conservation of the total number of bees which takes the form of a non-local, integral constraint on the optimal (most likely) gas density history $q(\mathbf{x},t)\geq 0$:
\begin{equation}\label{conservq}
 \int \displaylimits_{|\mathbf{x}|<L(t)}  q(\mathbf{x},t) \,d\mathbf{x} = 1 \quad \text{for all} \quad 0\leq t\leq T\,.
\end{equation}
To accommodate this constraint in the OFM formalism we introduce a Lagrangian multiplier $\lambda(t)$ (which in general depends on time) and add the term
\begin{equation}\label{lambdaterm}
\int_0^T dt\,\lambda(t) \int \displaylimits_{|\mathbf{x}|<L(t)}  q(\mathbf{x},t) \,d\mathbf{x}
\end{equation}
to the Hamiltonian of the classical field-theory \cite{EK,MS,MSK}.  Note that, in this form, the additional term (\ref{lambdaterm}) holds even for a more general constraint, when the total number of particles is a specified function of time, \textit{cf.} Ref. \cite{SmithMV}.

With the account of this additional term, the OFM equations \cite{EK,MS,MSK} become
\begin{eqnarray}
  \partial_t q &=& \frac{\delta H}{\delta p} = q e^p +\nabla \cdot \left(\nabla q-2 q \nabla p\right)\,, \label{q1} \\
  \partial_t p &=& -\frac{\delta H}{\delta q} = -\left(e^p-1\right) - \nabla^2 p- (\nabla p)^2 -\lambda(t)\,. \label{p1}
\end{eqnarray}
Here
\begin{equation}
\label{Hamiltonian}
H=H[q(\mathbf{x},t),p(\mathbf{x},t,\lambda(t))]= \int d\mathbf{x}\,\mathcal{H} (q,p,\lambda)
\end{equation}
is the constrained Hamiltonian,
\begin{equation}\label{Ham}
\mathcal{H}(q,p,\lambda) = \mathcal{H}_0(q,p) +\lambda(t) q
\end{equation}
is the density of the constrained Hamiltonian, and
\begin{equation}
\label{H0am}
\mathcal{H}_0(q,p) =(e^p-1) q -\nabla q\cdot \nabla p
+q \left(\nabla p\right)^2
\end{equation}
is the density of the unconstrained Hamiltonian \cite{EK,MS,MSK}.  The boundary conditions on the absorbing wall (or on the two absorbing walls for $d=1$) are \cite{bd,MFT,shpiel,main,b,fullabsorb}.
\begin{equation}\label{absorbingwall}
q(|\mathbf{x}|=L(t)) = p(|\mathbf{x}|=L(t)) = 0\,.
\end{equation}
We also have to specify some boundary conditions in time: at $t=0$ and $t=T$. These depend on whether we deal with a deterministic or fluctuating initial condition at $t=0$, and on whether we specify the whole density function $q(x,t=T)$ or only the size of its compact support $L(t=T)$.

The probability distribution $\mathcal{P}(\ell,N,T)$, can be found, up to a preexponential factor, from the relation
\begin{equation}\label{probanonstat}
-\ln \mathcal{P}(\ell,N,T) \simeq N S(\ell,T)\,,
\end{equation}
where
\begin{equation}\label{actionmostgeneral}
S(\ell,T) = \int_0^T dt \int \displaylimits_{|\mathbf{x}|<L(t)}d\mathbf{x}\, \left(p \partial_t q -\mathcal{H}_0\right)
\end{equation}
is the action per particle. Plugging Eqs.~(\ref{q1}) and (\ref{H0am}) into Eq.~(\ref{actionmostgeneral}), we obtain $S$ in terms of an integral along the optimal trajectory:
\begin{equation}\label{actionnonstat}
\!\!S(\ell,T)=\!\!\int_0^T dt \!\! \!\!\!\!\int \displaylimits_{|\mathbf{x}|<L(t)}\!\! \!\!\!\!d\mathbf{x} \left[q \left(p e^p -e^p+1\right) + q (\nabla p)^2 \right]\,.
\end{equation}
%

Let us discuss the physical meaning of the OFM equations. The term $(e^p-1)q$ in the unconstrained Hamiltonian ~(\ref{H0am}) describes the fluctuating branching reaction.  It appears already in zero spatial dimension. The factor $e^p$, coming from the random character of branching,  modifies the effective branching rate: it enhances the branching for $p>0$, and suppresses it for $p<0$. The other two terms in Eq.~(\ref{H0am}) are familiar from the macroscopic fluctuation theory of a gas of non-interacting random walkers \cite{MFT}. In particular,  the term $q (\nabla p)^2$  describes the fluctuational contribution to the particle flux, coming from the stochastic character of Brownian motion. The solution for $q(\mathbf{x},t)$  describes the history of the optimal density configuration of the bees. In its turn, the solution for $p(\mathbf{x},t)$ describes the history of the optimal realization of the noise in the system. These $q$ and $p$ are  \emph{deterministic} time-dependent fields which give a dominant contribution to the specified large deviation of the system. In the absence of fluctuations we have $p=0$ and $\lambda(t)=0$. In this case Eq.~(\ref{p1}) is obeyed trivially, and Eq.~(\ref{q1}) coincides with the deterministic equation~(\ref{deterministicd}).

For arbitrary $T$ the OFM problem, described above, is both complicated and non-universal: it is intrinsically time-dependent, and the solution strongly depends on the initial and final conditions.  Fortunately, the problem becomes much simpler in the limit of very large $T$. Here it is natural to assume that the optimal gas density $q(x,t)$ and the momentum density $p(x,t)$  are stationary, $L(t)$ is \emph{equal} to the specified $\ell$, and the Lagrange multiplier $\lambda(t)$ is constant. The stationarity holds for most of the time interval $0<t<T$ except for short non-universal transients close to $t=0$ and $t=T$ which contribute to the action only at a subleading order, and which will be ignored in the following.  The stationarity assumption leads to the\emph{ steady-state} OFM equations
\begin{eqnarray}
   q e^p +\nabla \cdot \left(\nabla q-2 q \nabla p\right)&=& 0, \label{eqq1} \\
   -\left(e^p-1\right) - \nabla^2 p- (\nabla p)^2 &=& \lambda\,, \label{eqp1}
\end{eqnarray}
whereas Eqs.~(\ref{probanonstat}) and (\ref{actionnonstat}) yield
\begin{equation}\label{Sandr}
S(\ell,T) = T R_d(\ell)
\end{equation}
with the rate function
\begin{equation}\label{action}
\!\!R_d(\ell)= \int\displaylimits_{|\mathbf{x}|<\ell}  d\mathbf{x} \left[q \left(p e^p -e^p+1\right) + q (\nabla p)^2\right]\,.
\end{equation}
The initial and final conditions for $q$ become irrelevant in the steady-state solution. Finally, if there are multiple stationary solutions, the one with the minimal action -- hence, the minimal $R_d(\ell)$ -- must be selected. Equations~(\ref{probanonstat}) and~(\ref{Sandr}) lead to the announced large-deviation behavior (\ref{LDform}) of the probability density $\mathcal{P}$.

A further simplification arises when we go over from $q$ and $p$ to the Cole-Hopf canonical variables $Q=q e^{-p}$ and $P=e^p-1$. Equations~(\ref{eqq1}) and (\ref{eqp1}) become
\begin{eqnarray}
\nabla^2 Q + (2P+1+\lambda) Q &=& 0, \label{eqQ2} \\
\nabla^2 P + (P+1)(P+\lambda)&=&  0\,.\label{eqP2}
\end{eqnarray}
Note that $Q(\mathbf{x})\geq 0$, whereas $P(\mathbf{x})$ can vary from $-1$ to $+\infty$. Importantly, Eq.~(\ref{eqP2}) is decoupled from Eq.~(\ref{eqQ2}). Also, for a given $P(\mathbf{x})$, Eq.~(\ref{eqQ2}) is a linear and homogeneous equation for $Q(\mathbf{x})$.

In the Hopf-Cole variables the boundary conditions for Eqs.~(\ref{eqQ2}) and (\ref{eqP2}) are
\begin{equation}\label{BCsedge2}
Q(|\mathbf{x}|=\ell) = P(|\mathbf{x}|=\ell) = 0\,,
\end{equation}
the mass conservation~(\ref{conservq}) reads
\begin{equation}\label{conservQ}
\int\displaylimits_{|\mathbf{x}|<\ell}  Q(\mathbf{x}) [P(\mathbf{x})+1]\, d\mathbf{x} =1\,.
\end{equation}
and the rate function $R_d(\ell)$ becomes \cite{MS}
\begin{equation}\label{snewvar}
R_d(\ell) = - \int\displaylimits_{|\mathbf{x}|<\ell}  d\mathbf{x} \left[Q(P+P^2) - \nabla P \cdot \nabla Q\right]\,.
\end{equation}
Applying the first Green's identity and using the boundary condition $Q(|\mathbf{x}|=\ell)=0$ and Eq.~(\ref{eqP2}), we can transform Eq.~(\ref{snewvar}) to
\begin{equation}\label{actionrate}
R_d(\ell)=\lambda  \int\displaylimits_{|\mathbf{x}|<\ell}  d\mathbf{x}\, Q(P+1)\,,
\end{equation}
which, by virtue of the integral constraint~(\ref{conservQ}), brings us to the remarkably simple result
\begin{equation}\label{action3}
R_d(\ell) = \lambda\,.
\end{equation}
That is, the calculation of the rate function $R_d(\lambda)$ only requires  to express the Lagrange multiplier $\lambda$ through $\ell$. Equation~(\ref{action3}) implies that the solution exists only for $\lambda\geq 0$, as we indeed find here.

Although Eqs.~(\ref{eqQ2}) and (\ref{eqP2}) are written in the general $d$-dimensional form, we have assumed that the optimal solution is spherically symmetric:  $Q=Q(r)$ and $P=P(r)$, where $r$ is the radial coordinate. Therefore, only the radial part  $\nabla_r^2$ of the Laplace operators actually appears in Eqs.~(\ref{eqQ2}) and (\ref{eqP2}). Correspondingly, the spatial integration in Eq.~(\ref{conservQ}) is the following:
\begin{equation}\label{conservQr}
\Omega_d \int_0^{\ell} dr \,r^{d-1} Q(r) [P(r)+1]\,  =1\,,
\end{equation}
where $\Omega_d = 2 \pi^{d/2}/\Gamma(d/2)$ is the surface area of the
$d$-dimensional unit sphere, and $\Gamma(z)$ is the gamma function. For $d=1$
we have
\begin{equation}\label{conservQ1d}
\int_{-\ell}^{\ell} dx\, Q(x) [P(x)+1]\,  =1\,,
\end{equation}
and $Q(x)$ and $P(x)$ are even functions:
\begin{equation}\label{BCsymmetric}
Q^\prime(0)=P^\prime(0)=0\,.
\end{equation}
Here and in the following the primes stand for the $x$- (or $r$-) derivatives.

Equations~(\ref{eqQ2}) and (\ref{eqP2}) possess an important  conservation law which follows from their Hamiltonian (or, alternatively, Lagrangian) character:
\begin{equation}\label{conslawlambdad}
\left[Q^{\prime}P^{\prime} +Q(P+1)(P+\lambda)\right]^{\prime}+\frac{2(d-1)}{r}  Q^{\prime}P^{\prime} =0\,.
\end{equation}
One way of  deriving this conservation law is by using a direct analogy with the energy-momentum tensor of the classical field theory \cite{LLfieldtheory}. In this interpretation Eq.~(\ref{conslawlambdad}) expresses the vanishing divergence of the energy-momentum tensor in the presence of spherical symmetry.

The case of $d=1$ is special, because in this case the conservation law (\ref{conslawlambdad}) does not depend explicitly on the coordinate,
and we obtain \cite{MSK,MSFKPP}:
\begin{equation}\label{conslawlambda}
Q^{\prime} P^{\prime}+Q(P+1)(P+\lambda)=W = \text{const}\,.
\end{equation}
Furthermore, in the case of $d=1$, and only in this case, Eq.~(\ref{eqP2}) has its own conservation law:
\begin{equation}\label{Econstlambda}
\frac{1}{2} \left(P^{\prime}\right)^2+V\left(P,\lambda\right) =E = \text{const} \,.
\end{equation}
The cubic potential
\begin{equation}\label{V}
V(P,\lambda) = \frac{1}{3}P^3+\frac{1+\lambda}{2}P^2+\lambda P
\end{equation}
has extrema at $P=-1$ and $P=-\lambda$. The two conservation laws, Eqs. ~(\ref{conslawlambda}) and (\ref{Econstlambda}), make the one-dimensional case exactly integrable, see Sec. \ref{1dcase}. In particular, it follows that the optimal density profile in one dimension is mirror symmetric with respect to the origin: $x\leftrightarrow -x$.

For all $d$, the spherically symmetric stationary problem in the Hopf-Cole variables can be conveniently solved numerically. Since Eq.~(\ref{eqQ2})
is linear and homogeneous, and Eq.~(\ref{eqP2}) is not coupled to Eq.~(\ref{eqQ2}), we can solve Eqs.~(\ref{eqQ2}) and~(\ref{eqP2})
by setting an arbitrary nonzero boundary condition for $Q(r=0)$, for example $Q(r=0)=1$, and ultimately normalize the solution to unity
by using Eq.~(\ref{conservQr}). Specifying $\lambda$ and using $P(r=0)$ as a shooting parameter, we demand that the first zeros of $Q(r)$ and $P(r)$ coincide to a given accuracy, which gives us $\ell$ and the optimal profiles of $Q(r)$ and $P(r)$. See Fig. \ref{fig3Dlong} below for an example of such a numerical solution for $d=3$.

\section{Asymptotics of the rate function}
\label{threeasymp}

There are three asymptotic regimes, where the rate function $R_d(\lambda)$ can be calculated analytically in any dimension.

\subsection{$|\ell-\ell_0|\ll \ell_0$}
\label{ellclose}

Here $\lambda\ll 1$ is a small parameter, and the solutions of Eqs.~(\ref{eqQ2}) and (\ref{eqP2}) can be found via the perturbative ansatz
\begin{eqnarray}
P(r)&=&\sqrt{\lambda} \left[P_0(r)+\sqrt{\lambda} P_1(r) +\lambda P_2(r) + \dots \right] ,\label{LTP}\\
Q(r)&=&Q_0(r)+\sqrt{\lambda}\, Q_1(r)+\lambda Q_2(r)+\dots
\, ,\label{LTQ}
\end{eqnarray}
where the functions $P_0$, $P_1$, $Q_0$, $Q_1$, $\dots$, are $O(1)$. In the leading order the ansatz~(\ref{LTP}) and~(\ref{LTQ}) leads to two identical Helmholtz equations
\begin{eqnarray}
\nabla^2_r P_0 + P_0 &=& 0\,,\label{eqP0} \\
\nabla^2_r Q_0 + Q_0 &=& 0\,,
\label{eqQ0}
\end{eqnarray}
which coincide with Eq.~(\ref{Helmholtz}).  In the subleading order we obtain
\begin{eqnarray}
\nabla^2_r P_1 + P_1  &=& -1-P_0^2
\, ,\label{eqP1} \\
\nabla^2_r Q_1 + Q_1&=& -2P_0 Q_0\,,\label{eqQ1}
\end{eqnarray}
\textit{etc.}
The solutions of Eqs.~(\ref{eqP0}) and~(\ref{eqQ0}) can be written as
\begin{equation}\label{solP0Q0}
\frac{P_0(r)}{p_0} =\frac{Q_0(r)}{q_0} =\,2^{\frac{d}{2}-1} \Gamma
   \left(d/2\right) F(r)\,,
\end{equation}
where $p_0$ and $q_0>0$ are (yet unknown) constants, and $F(r)$ was defined in Eq.~(\ref{F}).
For $d=1$, $2$ and $3$ we obtain
\begin{numcases}
{\frac{P_0(r)}{p_0} =\frac{Q_0(r)}{q_0} =}\,\cos r, & $d=1$, \nonumber \\
 \,J_0(r) ,& $d=2$, \nonumber\\
     \frac{\,\sin r}{r},& $d=3$,    \label{d123}
\end{numcases}
respectively.  For any $p_0$ and $q_0$, the two functions $P_0(r)$ and $Q_0(r)$ have their first positive zeros at the same point $r=\ell_0$. The subleading-order equations~(\ref{eqP1}) and (\ref{eqQ1}) lift this degeneracy. The condition that the first positive zeros of $P(r)$ and $Q(r)$ coincide also in the subleading order yield a unique value of the constant $p_0$. Having found $P(r)$ in the subleading order, we can determine the rate function $R_d(\ell)=\lambda$ in terms of the small difference $\ell-\ell_0$.  The remaining constant $q_0$ can be determined from Eq.~(\ref{conservQr}), but it does not affect $\lambda$.

The calculation proceeds as follows. First, we solve the linear equations~(\ref{eqP1}) and (\ref{eqQ1}) with the boundary conditions $P_1(0)=P_1^{\prime}(0)=0$ and  $Q_1(0)=Q_1^{\prime}(0)=0$, respectively. The solutions can be written as
\begin{eqnarray}
  P_1(r) &=&\frac{\pi}{2} \int_0^r dy\,y^{d-1} \begin{vmatrix} F(r) & F(y) \\ G(r) & G(y) \end{vmatrix} (1+P_0(y)^2), \label{solP1} \\
  Q_1(r) &=&\pi  \int_0^r dy\,y^{d-1} \begin{vmatrix} F(r) & F(y) \\ G(r) & G(y) \end{vmatrix} P_0(y) Q_0(y). \label{solQ1}
\end{eqnarray}
where $F(r)$ is defined in Eq.~(\ref{F}),
\begin{equation}\label{G}
G(r) = \frac{Y_{\frac{d}{2}-1}(r)}{r^{\frac{d}{2}-1}}\,,
\end{equation}
and $Y$ is the Bessel function of the second kind. As $P_1$ and $Q_1$ give subleading contributions, it suffices to evaluate the integrals in Eqs.~(\ref{solP1}) and~(\ref{solQ1})  at $r=\ell_0$. This yields the special value of $p_0$ for which the zeros of $P$ and $Q$ coincide in the subleading order:
\begin{equation}\label{p0d}
p_0^2= \frac{\int_0^{\ell_0} dr\,r^{d-1} F(r)}{2^{d-2}[\Gamma(d/2)]^2\int_0^{\ell_0} dr\,r^{d-1} [F(r)]^3}\,.
\end{equation}
As a result,
\begin{numcases}
{{p_0^2} =} 3/2 , & $d=1$, \label{p01} \\
2.21501\dots\,,& $d=2$, \label{p02}\\
\frac{4\pi }{3 \,\text{Si}(\pi )-\text{Si}(3 \pi )} =3.23787\dots,& $d=3$. \label{p03}
\end{numcases}
where $\text{Si}(z) = \int_0^z dt\,\sin (t)/t$  is the sine integral function.  For $\ell < \ell_0$ we have $p_0>0$. Here the fluctuations enhance branching ($p>0$) and cause an inward particle flux ($\nabla_r p <0$). The case of $\ell>\ell_0$ corresponds to $p_0 < 0$. Here the branching is suppressed, and an outward particle flux is enhanced, by fluctuations.

With $p_0^2$ at hand, we Taylor expand $P_0(r)$ in a small vicinity of $r=\ell_0$, keeping only the term linear in $\ell-\ell_0$,
\begin{equation}\label{P0prime}
P_0(r) \simeq P_0^{\prime}(\ell_0) (\ell- \ell_0)\,,
\end{equation}
and express $\lambda$ via the small shift $\Delta \ell = \ell-\ell_0$ of the first zero of $P(r) \simeq P_0(r)+\sqrt{\lambda} P_1(r)$. In this way we obtain the asymptotic of the rate function $R_d(\ell) = \lambda$, announced in Eq.~(\ref{wnlineartheory}): $R_d(\ell) \simeq \alpha_d (\Delta \ell)^2$
with
\begin{eqnarray}
 \!\!\!\alpha_d  &=& 
 \!\frac{\left[P_0^{\prime}(\ell_0)\right]^2}{\left[\pi G(\ell_0) \int_0^{\ell_0} dr\,r^{d-1} F(r)\right]^2} \nonumber \\
  \!\!\!&=& \! \frac{\left[F^{\prime}(\ell_0)\right]^2}{[\pi G(\ell_0)]^2 \int_0^{\ell_0} dr\,r^{d-1} F(r) \int_0^{\ell_0} dr\,r^{d-1} [F(r)]^3},
  \label{alphad}
\end{eqnarray}
where we used Eqs.~(\ref{solP0Q0}) and~(\ref{p0d}). Equation~(\ref{alphad}) gives $\alpha_1=3/8$, $\alpha_2 = 0.14924\dots$, and
\begin{equation}\label{alpha3}
\alpha_3=
\frac{1}{\pi [3 \,\text{Si}(\pi )-  \text{Si}(3 \pi
   )]} =0.08201\dots \,.
\end{equation}
The quadratic dependence of $R_d$ on $\Delta \ell$  describes a Gaussian asymptotic of the distribution $\mathcal{P}(\ell,N,T)$ at $\ell$ close to the expected value $\ell = \ell_0$. 

\subsection{$\ell\ll \ell_0$}
\label{smallell}

Here the fluctuations should work against a strong outward diffusion and also enhance the branching process considerably so as to make up for the particle losses at $r=\ell\ll \ell_0$. The optimal balance between these two effects is determined by Eqs.~(\ref{eqQ2}) and~(\ref{eqP2}) at $\lambda\gg 1$. As the first zeros of $Q$ and $P$ must coincide (at $r=\ell$), we see that $P$ (which is positive and large here) must scale as $\sqrt{\lambda}$, whereas $\ell$ must scale as $1/\sqrt{\lambda}$.  This leads us to the large-$\lambda$ perturbative ansatz
\begin{eqnarray}
P(r)&=&\sqrt{\lambda}\left[P_0 (\bar{r})+\frac{1}{\sqrt{\lambda}} P_1(\bar{r}) + \frac{1}{\lambda} P_2(\bar{r})+\dots \right]
 ,\label{largeP}\\
Q(r)&=&q_0\left[Q_0(\bar{r})+\frac{1}{\sqrt{\lambda}}\, Q_1(\bar{r})+ \frac{1}{\lambda} Q_2(\bar{r})+\dots\right],
 \label{largeQ}
\end{eqnarray}
where $\bar{r}=r\sqrt{\lambda}$. The functions $P_0$, $P_1$, $Q_0$, $Q_1$, $\dots$, are again $O(1)$,  whereas $q_0$ must scale as $\lambda^{(d-1)/2}$ to comply with the normalization condition. This ansatz leads to the same pair of Helmholtz equations
\begin{eqnarray}
\nabla^2_{\bar{r}} P_0 + P_0 &=& 0\,,\label{eqP0large} \\
\nabla^2_{\bar{r}} Q_0 + Q_0 &=& 0
\label{eqQ0large}
\end{eqnarray}
as in Sec.~\ref{ellclose}. Remarkably, it also yields the same subleading-order equations,
\begin{eqnarray}
\nabla^2_{\bar{r}} P_1 + P_1  &=& -1-P_0^2
\, ,\label{eqP1large} \\
\nabla^2_{\bar{r}} Q_1 + Q_1&=& -2P_0 Q_0\,,\label{eqQ1large}
\end{eqnarray}
as Eqs.~(\ref{eqP1}) and~(\ref{eqQ1}). Comparing Eqs.~(\ref{LTP}) and~(\ref{LTQ}) with Eqs.~(\ref{largeP}) and~(\ref{largeQ}), we see that, for the purpose of coincidence of the zeros of $P$ and $Q$,  there is an exact mapping between the two cases if we replace $\sqrt{\lambda}$ by $1/\sqrt{\lambda}$ and, in the case of $\ell \ll \ell_0$, measure the distance in the units of $1/\sqrt{\lambda}$. This unexpected mapping between the two (physically very different) regimes
allows us to immediately obtain the rate function  asymptotic
$R_d(\ell) = \lambda(\ell)$ for $\ell \ll \ell_0$ from the already found asymptotic~$\lambda=\alpha_d (\ell - \ell_0)^2$ for $|\ell-\ell_0|\ll \ell_0$. After a simple algebra we obtain, in the leading and subleading orders in $1/\ell\ll 1$, the expression announced in Eq.~(\ref{ellsmall}). In particular,
\begin{numcases}
{{R_d(\ell\ll \ell_0)} \simeq} \frac{\pi^2}{4\ell^2} - \sqrt{\frac{32}{3}} \,\frac{1}{\ell}, & \!\!\!\!$d=1$ \label{r1smalll} \\
\frac{5.7831\dots}{\ell^2} - \frac{5.1771\dots}{\ell},& \!\!\!\!$d=2$ \label{r2smalll}\\
\frac{\pi^2}{\ell^2} - \,\frac{6.9838\dots}{\ell},& \!\!\!\!$d=3$ \label{r3smalll}
\end{numcases}
The leading-order asymptotic $R_d(\ell\ll \ell_0) \simeq \ell_0^2/\ell^2$ coincides with the rate function, corresponding to the long-time survival probability of \emph{pure} Brownian motion (no branching) inside a $d$-dimensional sphere of radius $\ell$, see \textit{e.g.} Ref. \cite{Agranovetal}. This result is not unexpected: at very small $\ell$ the fluctuations mostly ``work" against diffusion which otherwise would rapidly spread out the swarm.
The subleading term, proportional to $1/\ell$, is much larger than $1$ and is therefore important. It is already affected by the branching process.

We can also determine the optimal stationary profile of the gas density $q(r) =Q(r)[1+P(r)]$, conditioned on $\ell\ll \ell_0$.
In most of the region $r<\ell$,  $q(r)$ can be approximately described by the leading-order solution
\begin{eqnarray}
  q(r) &\simeq & q_0 \sqrt{\lambda}\, Q_0(\sqrt{\lambda} r) \,P_0(\sqrt{\lambda} r) \nonumber \\
 &\simeq& C_d\frac{\ell_0^d}{\ell^d}\left[F\left(\frac{\ell_0 r}{\ell}\right)\right]^2 \,.
 \label{mostofregion}
\end{eqnarray}
where  the function $F(r)$ is defined in Eq.~(\ref{F}), $\lambda \equiv R_d(\ell)$ is given by Eqs.~(\ref{ellsmall}) and (\ref{r1smalll})-(\ref{r3smalll}),
and the numerical factor $C_d$ is determined by normalization to unity.
Equation~(\ref{mostofregion}) coincides with the optimal density profile, corresponding to the survival probability of pure Brownian motion \cite{Agranovetal}. Importantly, however, Eq.~(\ref{mostofregion}) breaks down close to $r=\ell$, where the solution is dominated by the term $q\simeq Q\simeq q_0 Q_0(\sqrt{\lambda}r)$. It is this term which determines (large) particle losses at $r=\ell$ which are compensated by the strongly enhanced branching in the bulk \cite{largeflux}.

\begin{figure}
\centering
\includegraphics[width=5cm]{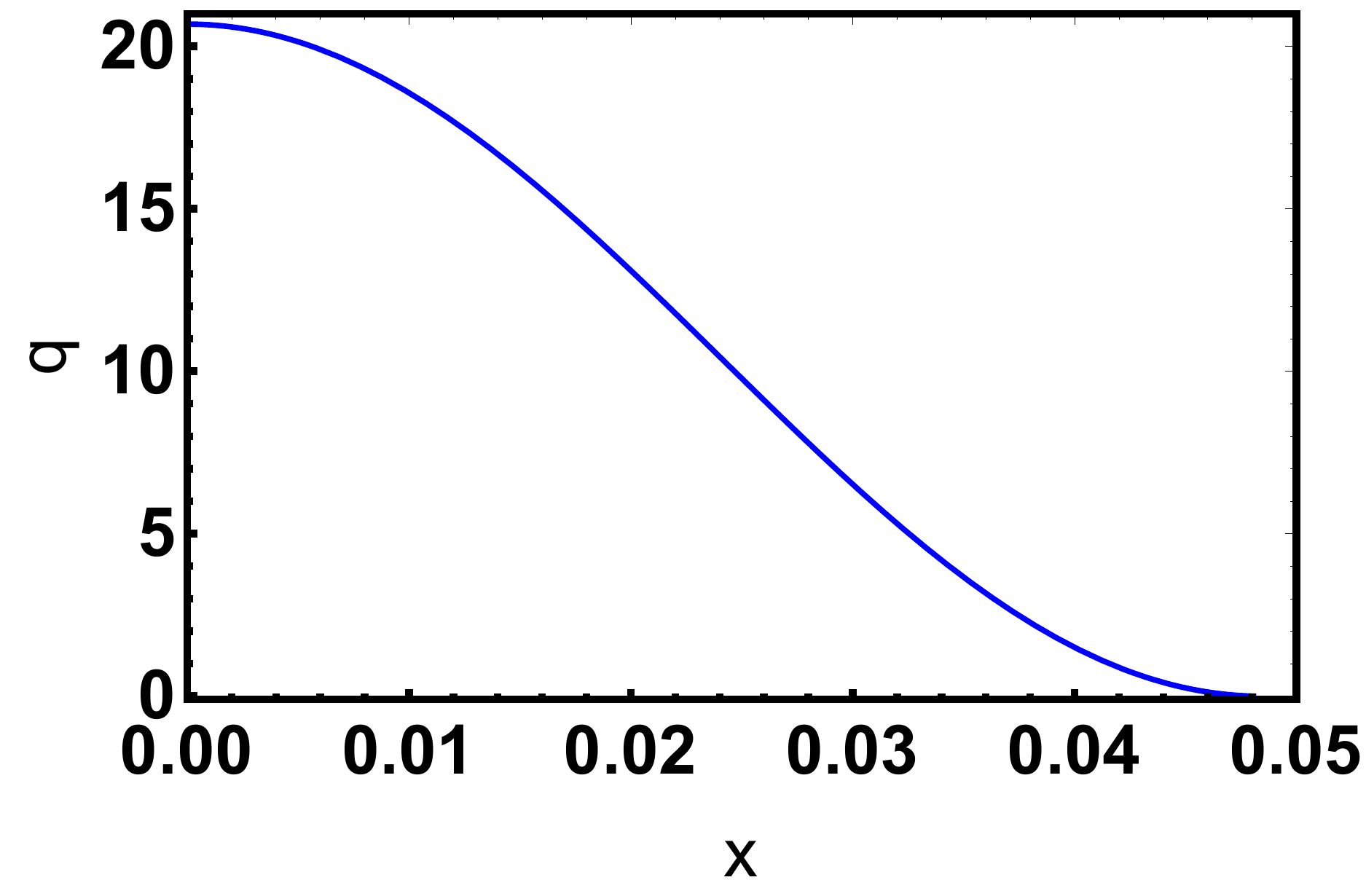}
\caption{The optimal density $q(x)$ for $d=1$ and $\ell \simeq 0.048 \ll \ell_0$. Here $\lambda=10^3$.}
\label{smallellq}
\end{figure}

As an example, Fig.~\ref{smallellq} shows the optimal density profile $q(\mathbf{x})$ for $d=1$ and $\ell =0.048\ll \ell_0$. Here $\lambda=10^3$.

\subsection{$\ell\gg \ell_0$}
\label{largeell}

In this regime the fluctuations must keep the swarm very large which requires a strong suppression of the branching process. A lower bound for $\mathcal{P}(\ell, N,T)$ [that is an upper bound for $R_d(\ell \gg \ell_0)$] can be obtained by assuming
that there are no branching events altogether during the whole time $T$. In this extreme scenario there are no particle losses, while the diffusion spread proceeds unhindered, that is deterministically. In this scenario the probability density
$\mathcal{P}^{\text{lower bound}} (\ell, N,T)=\exp(-NT)$
is independent of $\ell$ (and, in the physical units, independent of the diffusion constant), leading to the upper bound for $R_d$:
\begin{equation}\label{bound}
R_d^{\text{upper bound}} (\ell \gg \ell_0)= 1\,.
\end{equation}
As we will now show, the true optimal configuration at $\ell \gg \ell_0$ outperforms this simple upper bound by suppressing the branching and diffusion in most of the swarm, but allowing it in a close vicinity of $r=\ell$. Almost all of the bees in this regime are concentrated near the swarm boundary.

A complete suppression of the branching by fluctuations would require $p=-\infty$, that is $P=-1$. The true optimal trajectory is such that $P(r)$ stays very close to $-1$ on most of the interval $0<r<\ell$, and increases and reaches $P=0$ in the narrow boundary layer near $r=\ell$ with width $O(1)$. Let us first consider $d=1$.

\subsubsection{$d=1$}

Here there are two conservation laws~(\ref{conslawlambda}) and~(\ref{Econstlambda}). In the next section we explain how one can use them to solve the case of $d=1$ exactly for any $\ell$. Here we use them to find a simple \emph{limiting solution} $Q(x)$ and $P(x)$ which gives the leading-order asymptotic of the exact solution at $\ell \to \infty$. Because of the symmetry of the one-dimensional swarm with respect to $x=0$, we can limit ourselves to positive $x$, that is consider the right half of the swarm.

Plugging the exact boundary condition
$P^{\prime}(x=0)=0$ and the asymptotically exact boundary condition $P(x=0)= -1$ into Eq.~(\ref{conslawlambda}), we see that $W=0$. Then, because of the boundary condition $Q(x=\ell)=0$, the same equation yields $P^{\prime}(x=\ell)=0$ \cite{Qprime}.

\begin{figure}
\centering
\includegraphics[width=5cm]{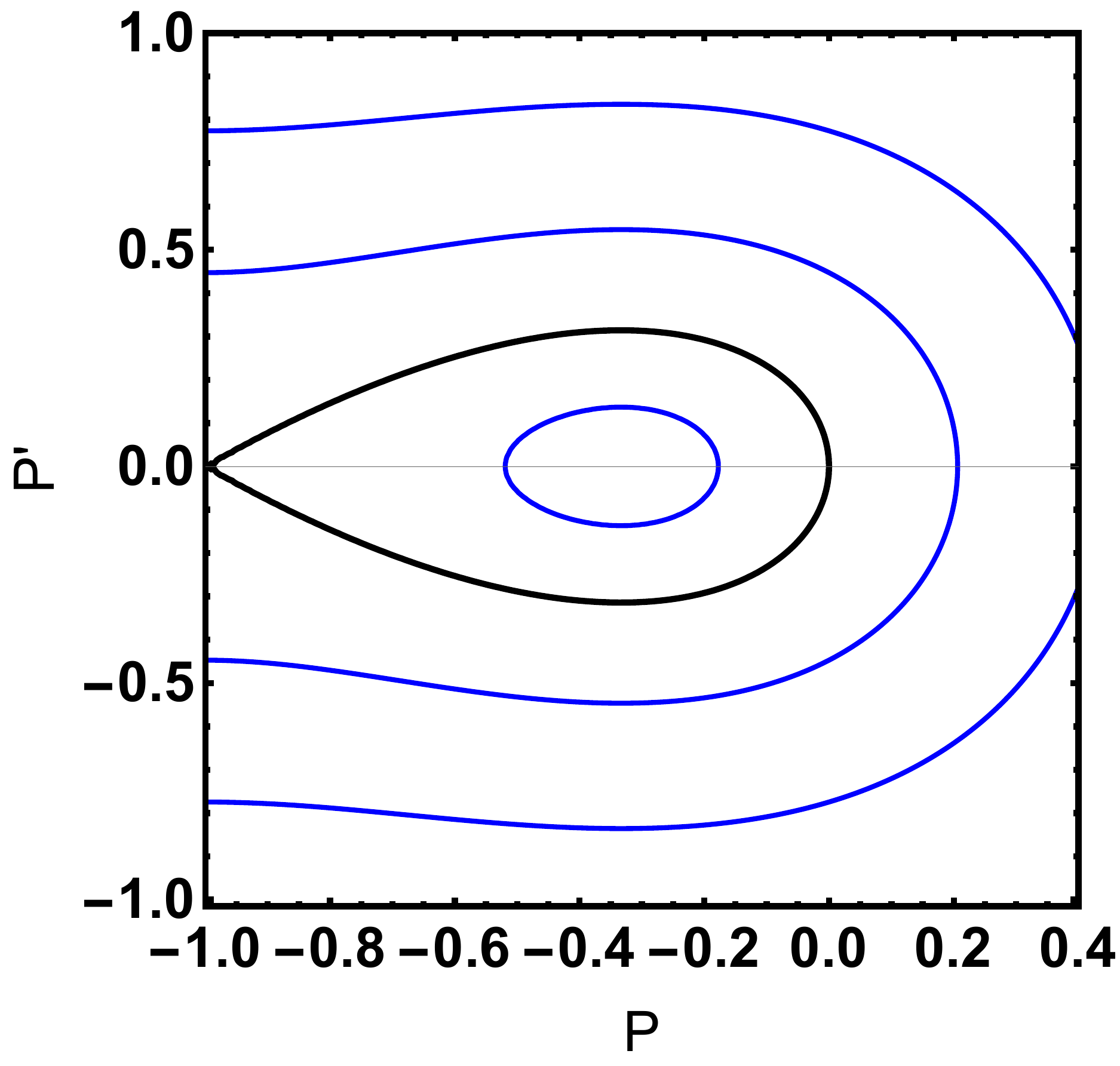}
\caption{The phase portrait $(P,P^{\prime})$, described by Eq.~(\ref{Econstlambda}), for $d=1$ and $\lambda=1/3$. The homoclinic trajectory $E=0$ corresponds to the limiting solution $P(x)$ for $\ell \to \infty$, see Eq.~(\ref{Psimple}).}
\label{phaseportraitlong}
\end{figure}

Now we turn to the second conservation law~(\ref{Econstlambda}). Using the equality $P^{\prime}(x=\ell)=0$, that we have just established, and the boundary condition $P(x=\ell)=0$, we obtain $E=0$. Then, using  $P^{\prime}(x=0)=0$ and $P(x=0)= -1$, we obtain $\lambda=1/3$.   This immediately leads us to the large-$\ell$ asymptotic of the rate function:
\begin{equation}\label{r1}
R_1 (\ell \gg \ell_0) \simeq \frac{1}{3}\,.
\end{equation}
This leading-order asymptotic is independent of $\ell$, and is three times smaller than the simple upper bound (\ref{bound}).  Note that the limiting solution for $P(x)$, with $\lambda=1/3$ and $E=0$, corresponds to a homoclinic trajectory on  the phase plane $(P,P^{\prime})$, see Fig. \ref{phaseportraitlong}.

\begin{figure}
\centering
\includegraphics[width=5cm]{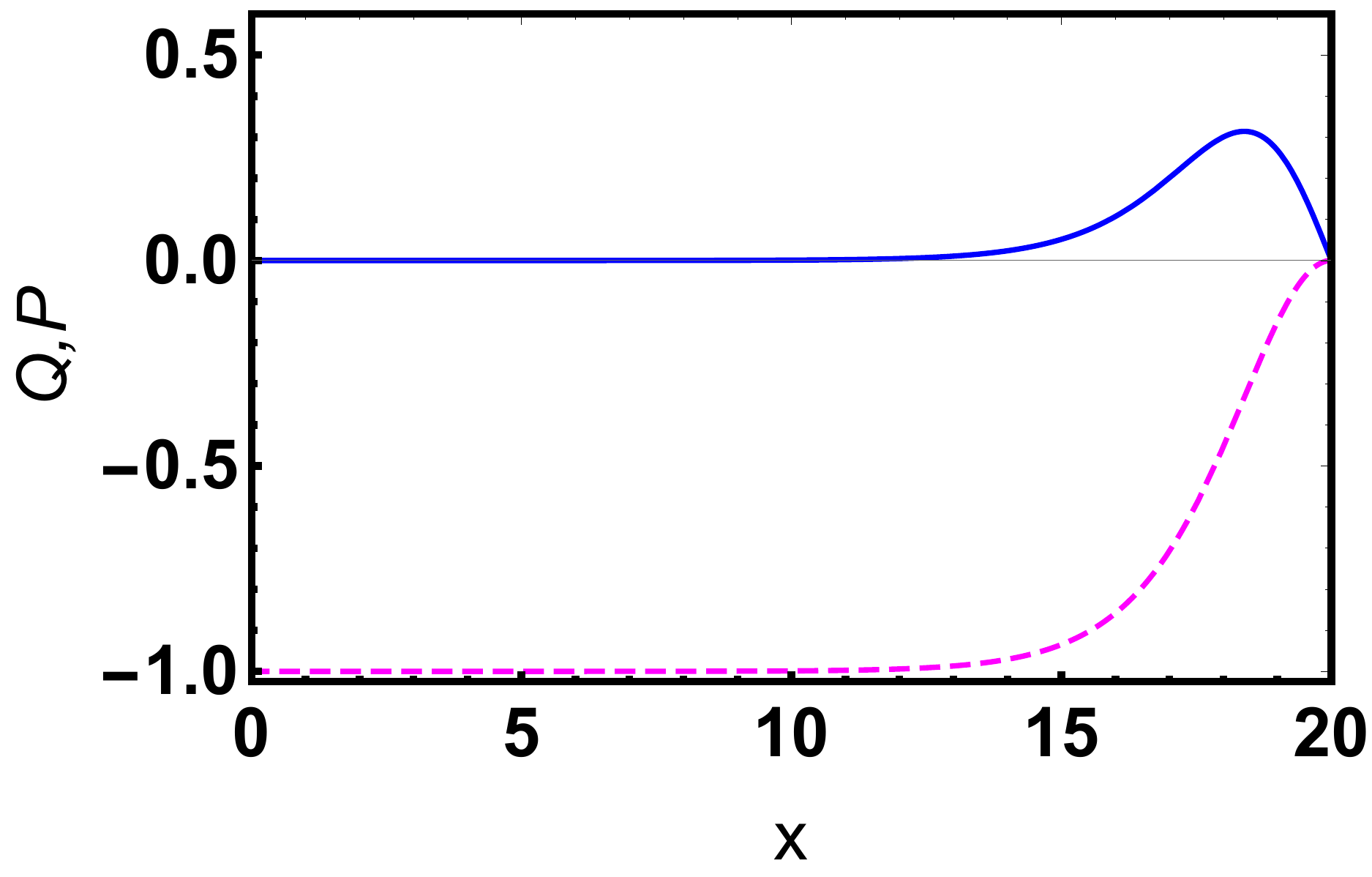}
\includegraphics[width=5cm]{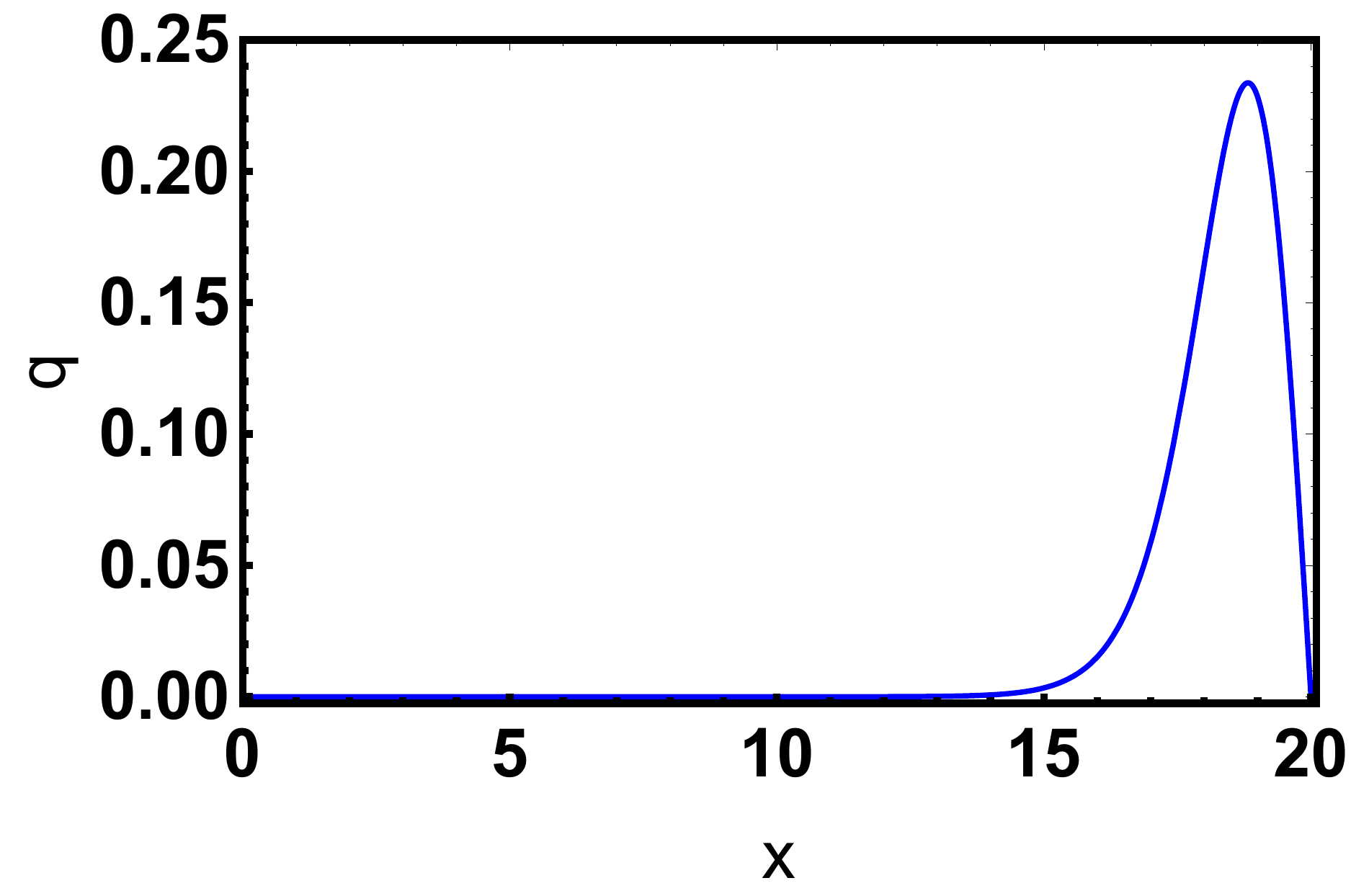}
\caption{The limiting optimal solution (\ref{Psimple})-(\ref{qsimple}) for $d=1$ and $\lambda=1/3$. Top: $Q(x)$ (solid line) and $P(x)$ (dashed line). Bottom: $q(x)$. In this figure we set $\ell =20$.}
\label{longanalyticsimple}
\end{figure}

In view of Eq.~(\ref{eqP2}), we can rewrite Eq.~(\ref{conslawlambda}) for the limiting solution as
\begin{equation}\label{conlawlambda2}
Q^{\prime} P^{\prime} -Q P^{\prime\prime} = 0\,.
\end{equation}
This homogeneous linear ODE  gives a simple relation between $Q(x)$ and $P(x)$ for the limiting solution: $Q(x) = k\,P^{\prime} (x)$, where $k>0$ is constant. In its turn,
\begin{equation}\label{qxpropk}
q(x) = [1+P(x)] Q(x) = k [1+P(x)] P^{\prime}(x)\,,
\end{equation}
and $k$ can be determined from the normalization condition:
\begin{equation}
\label{findingk}
k\int_{-\infty}^0  (1+P) P^{\prime}\, d\xi = \frac{k}{2} \left(1+P\right)^2\Big|_{\xi=-\infty}^{\xi=0} =\frac{1}{2}\,,
\end{equation}
so $k=1$. In fact, $P$ and $Q$, and  then $p$ and $q$, for the limiting solution can be found in an explicit and elementary form:
\begin{eqnarray}
  P(\xi) &=&  -1+\text{sech}^2\left(\frac{\xi}{\sqrt{6}}\right)\,, \label{Psimple}\\
  Q(\xi) &=& -\sqrt{\frac{2}{3}} \tanh \left(\frac{\xi}{\sqrt{6}}\right)
   \text{sech}^2\left(\frac{\xi}{\sqrt{6}}\right)\,, \label{Qsimple}\\
  p(\xi) &=& 2\,\ln \text{sech} \left(\frac{\xi}{\sqrt{6}}\right)\,, \label{psimple}\\
  q(\xi) &=& -\sqrt{\frac{2}{3}} \tanh \left(\frac{\xi}{\sqrt{6}}\right)
   \text{sech}^4\left(\frac{\xi}{\sqrt{6}}\right)\,, \label{qsimple}
\end{eqnarray}
where $\xi=x-\ell\leq 0$.  Figure \ref{longanalyticsimple} shows the resulting plots of $P(x)$ and $Q(x)$ (the top panel) and $q(x)$ (the bottom panel) for $\ell=20$.
As one can see, the optimal gas density, conditioned on $\ell \gg \ell_0$, is close to zero in most of the swarm. The bees are constantly produced in the boundary layer with width $O(1)$ near $x= \ell$, diffuse to the absorbing wall at $x =\ell$ and get absorbed. Furthermore, the optimal fluctuation suppresses the inward diffusion of the bees from the periphery. Indeed, using Eq.~(\ref{psimple}), we can see that, well outside of the boundary layer at $\xi=0$,  $p(\xi)$ behaves as $(2/\sqrt{6})\, \xi +\text{const}$. As a result, the $\nabla p$ term in Eq.~(\ref{eqq1}) produces a constant drift velocity directed outward. It is this outward drift which suppresses the inward particle diffusion.

\subsubsection{$d>1$}

Crucially, the leading-order asymptotic $R_d(\ell \gg \ell_0)=1/3$, obtained for $d=1$, is valid in all dimensions, as announced in Eq.~(\ref{elllarge}). This is because, as $\ell$ goes to infinity, the first-derivative terms  $\frac{d-1}{r}\,P^{\prime}(r)$ and $\frac{d-1}{r}\,Q^{\prime}(r)$
in the Laplace operators of Eqs.~(\ref{eqP1}) and (\ref{eqQ1}) become negligible compared with the second derivative terms in the region $\ell-r = O(1)$, where $P(r)+1$ and $Q(r)$ are localized.

\begin{figure}
\centering
\includegraphics[width=5cm]{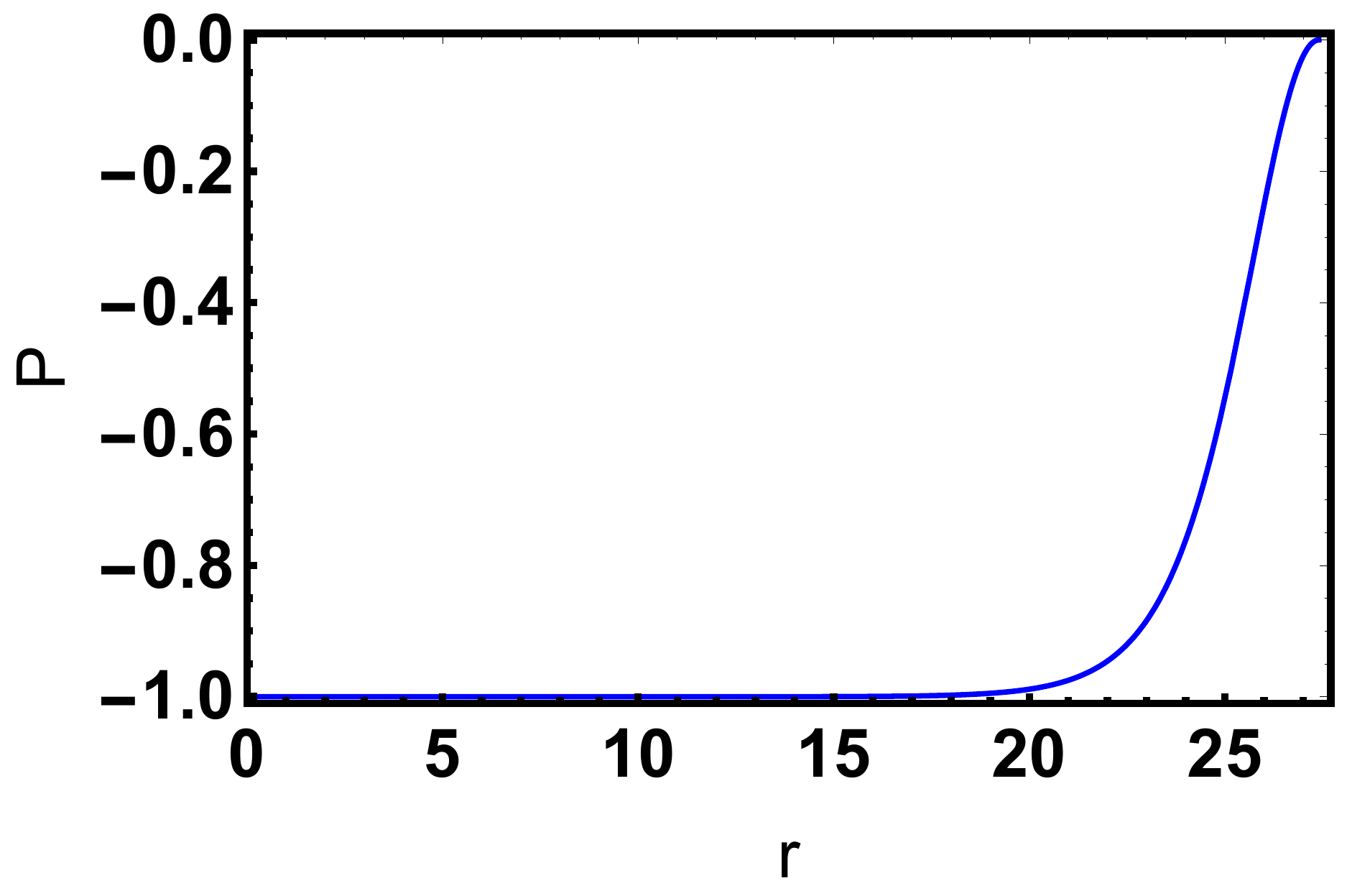}
\includegraphics[width=5cm]{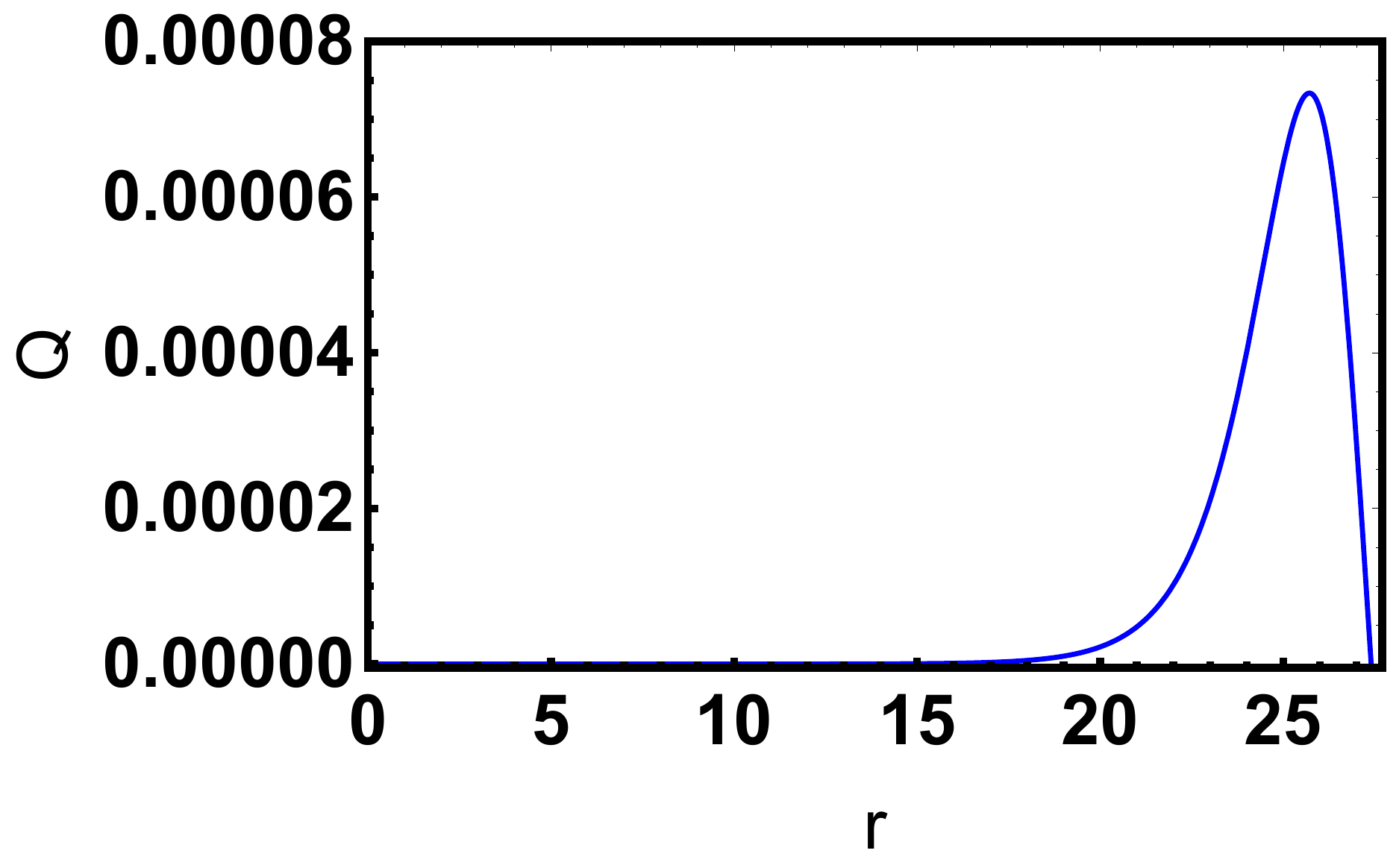}
\includegraphics[width=5cm]{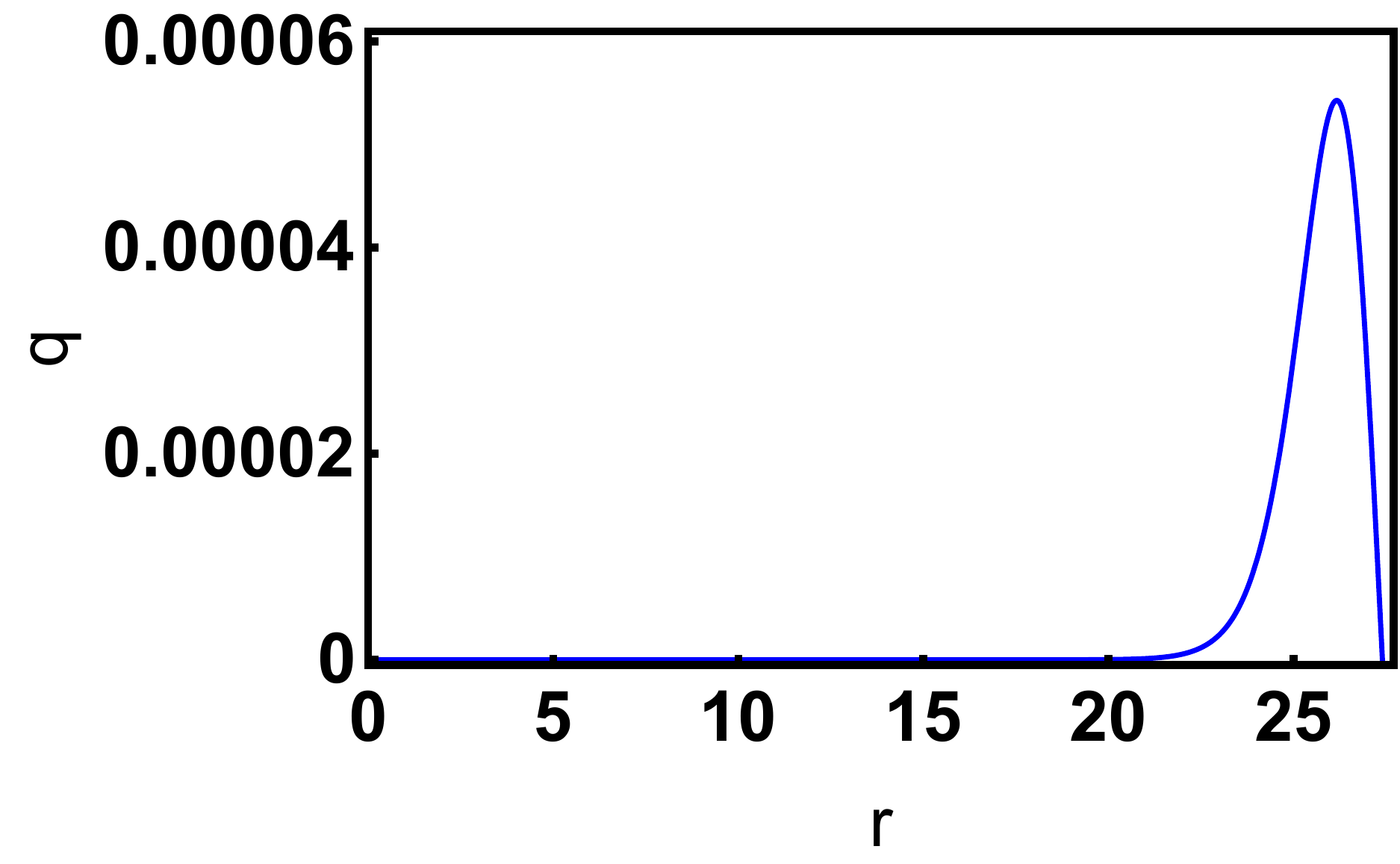}
\caption{The optimal solution found numerically by shooting for $d=3$ and $\lambda=0.3$. For this  $\lambda$ one obtains $\ell \simeq 27.4 \simeq 8.72\, \ell_0$.}
\label{fig3Dlong}
\end{figure}

As an example, Fig.~\ref{fig3Dlong} shows a numerical solution  for $d=3$ and $\lambda=0.3$, which corresponds to $\ell \simeq 27.4 \simeq 8.72 \ell_0$. As one can see, this optimal solution is qualitatively similar to that for $d=1$.
The branching is almost completely suppressed in the bulk of the swarm: $P(r)\simeq -1$. This, however,  does not lead to an action proportional to $\ell$, because  the gas density $q(r)$ is almost zero in the bulk. The branching and diffusion act only close to the boundary of the swarm  $r=\ell$, and the fluctuation-induced outward drift suppresses the inward diffusion of the bees.

An attentive reader could have noticed that an $\ell$-independent asymptotic of $R_d(\ell)$, and therefore of $\mathcal{P}(\ell,N,T)$, formally
leads to a divergent integral $\int_0^{\infty}  \mathcal{P}(\ell,N,T) \, d\ell$ and, therefore, to a non-normalizable distribution. The resolution of this paradox is the following. The asymptotic (\ref{elllarge}) assumes
stationarity of the optimal configuration. At fixed $\ell$, this assumption is valid only when the observation time $T$ is sufficiently large. If we instead fix $N$ and $T$ and keep increasing $\ell$, we will ultimately enter a non-stationary (and possibly non-spherically symmetric) regime, where $\mathcal{P}(\ell,N,T)$
does not have the large-deviation form~(\ref{LDform}), and where it is expected to rapidly fall off with a further increase of $\ell$, thus
resolving the non-normalizability paradox.

\section{$d=1$ is integrable}
\label{1dcase}

In one dimension, Eq.~(\ref{Econstlambda}) describes a Newtonian particle of unit mass with ``coordinate" $P$ moving in ``time" $x$ in the cubic potential $V(P,\lambda)$. The exact Newtonian trajectories $P(x)$ are given
in terms of the elliptic Jacobi functions or the elliptic Weierstrass functions, depending on whether the cubic equation $V(P,\lambda)=E$ has three real roots or one real root, respectively \cite{Schwalm}. The arbitrary shift of the solution $x \to x+\text{const}$ is eliminated by demanding that $P^{\prime}(0)=0$. The approximate asymptotic solutions for $P(x)$ that we obtained, for $d=1$, in Secs. \ref{ellclose} and \ref{smallell}, and in Sec. \ref{largeell}, arise in two opposite limits  when the elliptic functions reduce to elementary functions: to the cosine and to the squared hyperbolic secant, respectively \cite{Schwalm}.

\begin{figure}
  \centering
  \includegraphics[width=5cm]{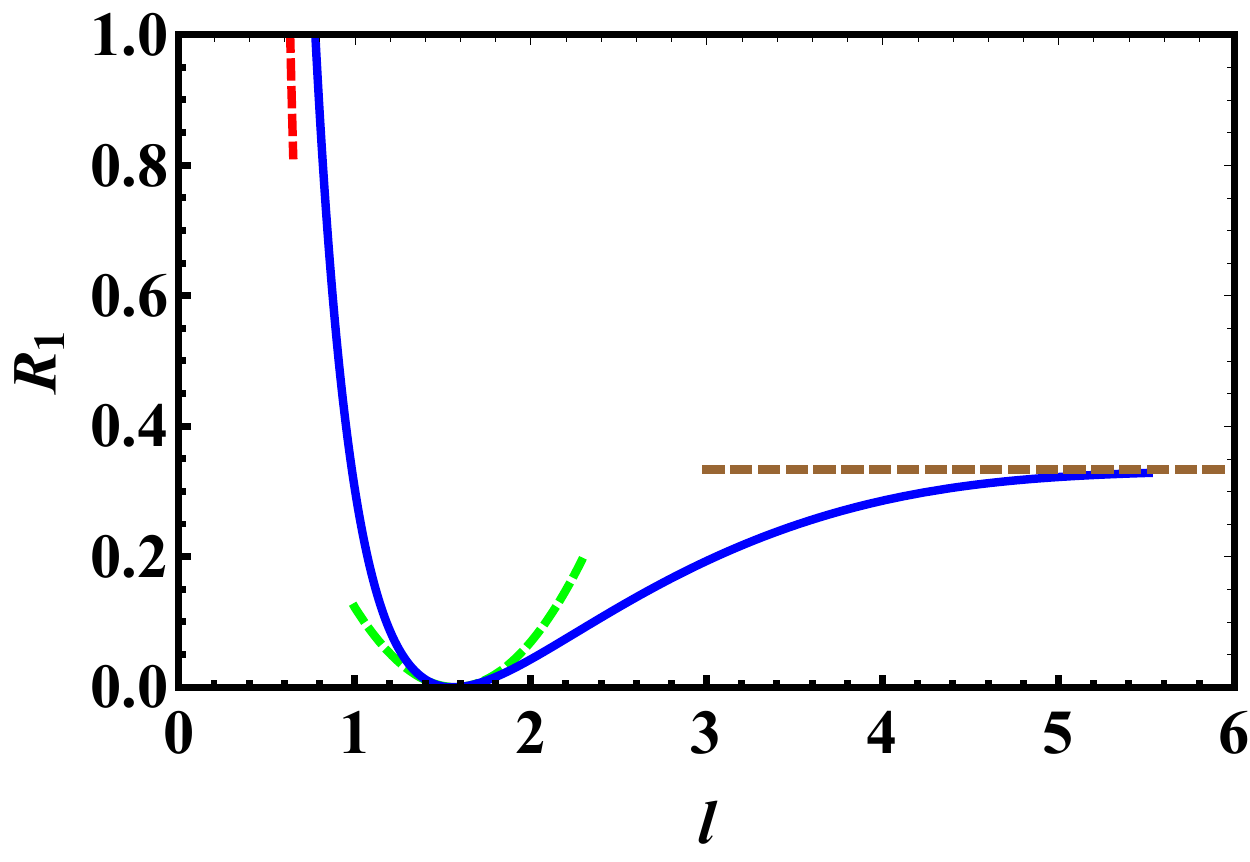}
  \includegraphics[width=5cm]{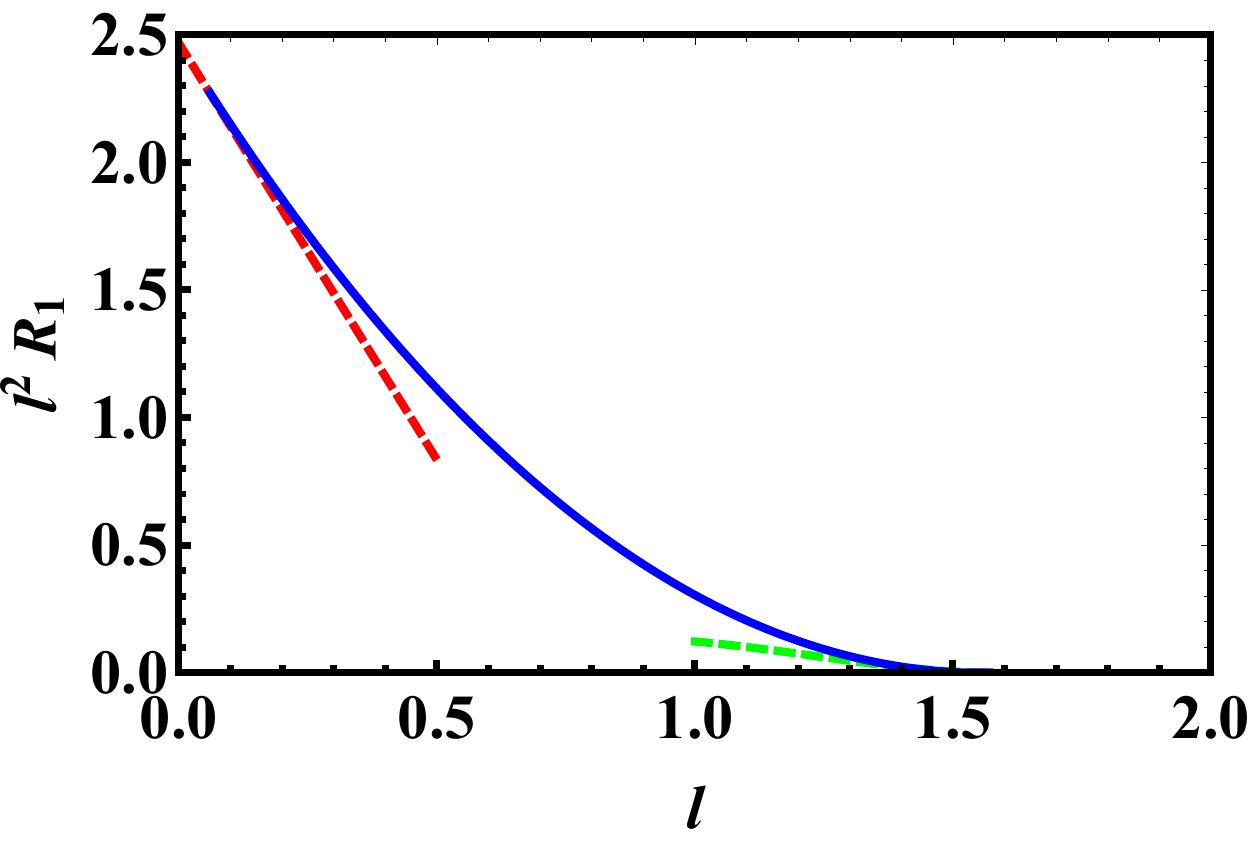}
  \caption{The rate function vs. $\ell$ in one dimension. Top: $R_1(\ell)=\lambda(\ell)$ (solid line) and the asymptotics~(\ref{wnlineartheory})-(\ref{elllarge}) (dashed lines). Bottom: $R_1 \ell^2$ vs. $\ell$ (solid line) alongside with the asymptotics ~(\ref{wnlineartheory}) and (\ref{ellsmall}),  multiplied by $\ell^2$ (dashed lines).}
  \label{1dfull}
\end{figure}

With the solution for $P(x)$ at hand, we turn to the second conservation law~(\ref{conslawlambda}) which, in view of Eq.~(\ref{eqP2}), can be recast as
\begin{equation}\label{firstorderode}
Q^{\prime} P^{\prime}-Q P^{\prime\prime}= W = \text{const}\,.
\end{equation}
For a given $P(x)$, Eq.~(\ref{firstorderode}) is a linear first-order ODE for $Q(x)$, which can be immediately solved. What is left is to fix three arbitrary constants  [$E,W$ and the additional constant entering the general solution of Eq.~(\ref{firstorderode})] and determine the value of $\ell$, where both $P(x)$ and $Q(x)$ become zeros. There are three conditions to obey: $Q^{\prime}(0)=0$ and $Q(\ell)=P(\ell) = 0$, which fix these three constants. These (quite cumbersome) algebraic conditions have to be solved numerically.  Finally, the normalization condition~(\ref{conservQ1d}) is used to fix the amplitude of $Q(x)$, although this last step is unnecessary for determining the rate function $R_1(\ell) = \lambda(\ell)$.

We implemented this scheme in full.  The explicit expressions [especially the ones involving $Q(x,E,W)$] are too bulky to be presented here. Therefore we only show, in Fig. \ref{1dfull}, the final results in the form of a plot of $R_1=R_1(\ell)=\lambda(\ell)$. Also shown are the three asymptotics (\ref{wnlineartheory})-(\ref{elllarge}) for $d=1$. The lower panel of Fig. \ref{1dfull} shows the product $ \ell^2 R_1(\ell)$ vs. $\ell$. Evident is a good agreement between the asymptotics and the exact results in the proper regions.

\section{Summary and Discussion}
\label{summary}

Here we studied persistent large deviations of the maximum distance of any of $N\gg 1$ Brownian bees from the origin in the limit of $T\gg 1$.  Assuming a spherically symmetric optimal density profile of the bees, centered at the origin, we determined the rate function $R_d(\ell)$, which characterizes these large deviations, analytically and numerically in different limits.   The optimal fluctuation method (OFM) was instrumental in
obtaining these results. In addition to the rate function itself, the OFM provides an illuminating insight into the most probable configurations of the swarm that dominate the probability density of the specified unusual swarm size. As we observed, these configurations are quite fascinating in the limit of unusually large swarms, $\ell \gg \ell_0$.

As it is evident from Fig.  \ref{1dfull}, there exists a value of $\ell=\ell_1>\ell_0$ such that, at $\ell >\ell_1$, the rate function $R_d(\ell)$ is nonconvex and, therefore, cannot be obtained from the G\"{a}rtner-Ellis theorem, see \textit{e.g.} Ref. \cite{Touchette2009}.

Our $N\gg 1$ and $T\gg 1$ results  rely on the assumption of spherical symmetry of the optimal density profile. Although this assumption is quite natural on the physical grounds, it is important to verify it. We have already done it in one dimension. We considered a \emph{non-symmetric} stationary swarm located at $-\ell_1<x<\ell_2$, where, without loosing generality, we can set $\ell_1<\ell_2$. The long-time probability density,
$-\ln \mathcal{P}(\ell_1,\ell_2,N,T)\simeq NT R(\ell_1,\ell_2)$, can again be found by solving a stationary OFM problem.
The boundary conditions at $x=\ell_2$ remain absorbing as before: $q(\ell_2) = p(\ell_2)=0$. But as there is no bee loss at $x=\ell_1$, the boundary condition here changes into a \emph{no-flux} condition: $q'-2q p'=0$.   When combined with the zero-density condition $q(\ell_1) = 0$, this leads to $p(\ell_1)=-\infty$. In the Hopf-Cole variables $Q$ and $P$ the boundary conditions at $x=\ell_1$ are $Q=0$ and $P=-1$. As this OFM problem is invariant to translations in $x$, the rate function $R(\ell_1,\ell_2)$ depends only on the sum $\ell_1+\ell_2$. As we observed numerically and analytically, the resulting rate function $R(\ell_1,\ell_2)$ is always larger than the one we calculated in the main part of the paper for $\ell_1=\ell_2$ and symmetric boundary condition. Moreover, the difference of the rates $R(\ell_1\to\ell_2,\ell_2)-R(\ell_2)$ is a finite positive number which depends on $\ell_2$. The conclusion is that the symmetric configuration is the most probable. It would be interesting to test the spherical-symmetry assumption in higher dimensions as well.

It would be also very interesting to evaluate the probability  of \emph{instantaneous} fluctuations of the swarm size in the steady state. The corresponding
probability density $\mathcal{P}(N,\ell)$ does not depend on time. Solving this challenging problem would be a considerable achievement.

\section*{ACKNOWLEDGMENTS}

We thank N. R. Smith and A. Vilenkin for useful discussions. The research of B.M. is supported by the Israel Science Foundation (grant No. 1499/20). The research of P.S. is supported by the project ``High Field Initiative" (CZ.02.1.01/0.0/0.0/15\_003/0000449) of the European Regional Development Fund.


\begin{thebibliography}{99}
\bibitem{rescaling} By rescaling time and the coordinate one can set the branching rate and the diffusion constant to unity.
\bibitem{largeflux} The particle flux $\mathbf{j} = -\nabla q+2 q \nabla p$ at $r=\ell$ is equal to $-\nabla q$, or to $-\nabla Q$ in the Hopf-Cole variables. Hence, it is determined, in the leading order, by the expression $-q_0\sqrt{\lambda} \,Q_0^\prime|_{\bar{r}=\ell_0} $. As a result, the particle flux scales as $(\ell_0/\ell)^d\gg 1$.
\bibitem{Qprime} $Q^{\prime}(x=\ell)$ must be nonzero. If it were zero alongside with $Q(x=\ell)$, $Q(x)$ would vanish identically.
\end{thebibliography}
\end{document}